\documentclass[reprint,aps,nofootinbib,amsmath,prd,twocolumn,showpacs,superscriptaddress,groupedaddress]{revtex4-1}

\usepackage{graphicx}  
\usepackage{dcolumn}   
\usepackage{bm}        
\usepackage{amssymb}   

\usepackage[utf8]{inputenc}
\usepackage{slashed}
\usepackage{epsfig}
\usepackage{amsmath,amsfonts,mathtools}
\usepackage{color}

\def\mtxt#1{\quad\hbox{{#1}}\quad}

\def\E{\mathrm{e}}
\def\D{\mathrm{d}}
\def\tr{\mathrm{Tr\,}}
\def\ha{\frac{1}{2}}

\def\beq{\begin{equation}}    
\def\eeq{\end{equation}}      

\def\bea{\begin{eqnarray}}
\def\eea{\end{eqnarray}}

\def\tr{\,\mbox{tr}\,}                  
\def\pa{\partial}                       

\def\be{\beta}
\def\ch{\chi}

\def\la{\lambda}

\def\Ga{\Gamma}

\def\Si{\Sigma}

\def\nn{\nonumber}

\begin{document}

\mbox{}

\title{
Renormalization group flows and fixed points\\
for a scalar field in curved space 
with nonminimal $F(\phi)R$ coupling
}

\author{Boris S.\ Merzlikin}
\email{merzlikin@tspu.edu.ru}
\affiliation{Tomsk State Pedagogical University, Tomsk, 634061, Russia}

\author{Ilya L.\ Shapiro}
\email{shapiro@fisica.ufjf.br}
\affiliation{Departamento de F\'{\i}sica, ICE, Universidade Federal de Juiz de Fora
Juiz de Fora,\\ CEP: 36036-330, MG, Brazil}
\affiliation{Tomsk State Pedagogical University, Tomsk, 634061}
\affiliation{Tomsk State University, Tomsk, 634050, Russia}

\author{Andreas Wipf}
\email{wipf@tpi.uni-jena.de}
\affiliation{Friedrich-Schiller-Universit{\"a}t,
Theoretisch-Physikalisches-Institut,\\
Max-Wien-Platz 1,07743, Jena, Germany}

\author{Omar Zanusso}
\email{omar.zanusso@uni-jena.de}
\affiliation{Friedrich-Schiller-Universit{\"a}t,
Theoretisch-Physikalisches-Institut,\\
Max-Wien-Platz 1,07743, Jena, Germany}


\begin{abstract}
Using covariant methods, we construct and explore the Wetterich
equation for a non-minimal coupling $F(\phi)R$ of a
quantized scalar field to the Ricci scalar of a prescribed curved space.
This includes the often considered non-minimal coupling
$\xi \phi^2 R$ as a special case. We consider the
truncations without and with scale- and field-dependent
wave function renormalization in dimensions between four and two.
Thereby the main emphasis is on analytic and numerical solutions of the
fixed point equations and the behavior in the vicinity
of the corresponding fixed points. We determine the non-minimal
coupling in the symmetric and spontaneously broken phases
with vanishing and non-vanishing average fields, respectively.
Using functional perturbative renormalization group methods,
we discuss the leading universal contributions to the RG flow
below the upper critical dimension $d=4$.
\end{abstract}

 \maketitle

\section{Introduction}

The renormalization group (RG) method is a flexible and
powerful tool to study Quantum Field Theory in curved space-time. The
traditional perturbative formulation was initiated in the papers
 \cite{nelspan82,buch84,Toms83} and is reviewed in \cite{book}.
Unfortunately, this formulation is essentially
restricted to the Minimal Subtraction scheme of renormalization which hinders
its applicability to infrared scales. Hence
its applications to physical situations such as inflation or
acceleration of the present-day universe require a great amount of
phenomenological settings. This means that in many cases we are
unable to derive the most relevant part of the quantum corrections,
and therefore have to rely on general arguments based on
covariance and dimension (see, e.g., \cite{PoImpo} for a
review).

The quantized scalar field coupled to gravity has always attracted
a certain amount of interest. Recently this topic has again become
interesting due to the role that gravitational effects might
have on the Higgs decay, which could explain the stability of the Higgs
potential at high energies (see, e.g., \cite{Higgs_decay} for
the latest situation and further references). 
Another important motivation is
the growing interest on the effective potential of the Higgs field
itself and its  application to cosmology. Despite the limitations of the standard
perturbative RG-methods in curved space-time,
such a potential can be useful for consistently describing inflation
\cite{Shaposh-08,BKS-2008}. In particular, it is known that
the first- and second-loop corrections to the potential enable
us to impose restrictions on the mass of the Higgs particle
\cite{Shaposh-08,BKS-2008}. This means that the Higgs inflation,
taking the opportune RG corrections into account,
can provide falsifiable predictions for observational cosmology
\cite{Tronconi:2017wps}. The importance of the higher-loop corrections
and the sensitivity of the results to infrared effects  indicate that it may be
worth employing a non-perturbative method of renormalization, especially to
investigate the non-minimal coupling between the Higgs field and the
scalar curvature.

Some well-known non-perturbative methods can be applied to
curved space-time (see the reviews \cite{zinn,Wipf}). Among these
methods we include the functional renormalization group (FRG) approach \cite{FRG-review}
which has been much developed over the past decade, but
has still been very little used to study quantum field theories in a
\emph{curved background space}. Exceptions are \cite{benedetti}, in which the
critical behavior of scalar field theories on spherical and hyperbolic
spaces in the local potential approximation has been studied, and
\cite{Fr}, in which the symmetry restoration in de Sitter space has
been investigated within the same approximation. Most papers
adopting FRG methods to investigate the renormalization of matter
fields in curved spaces considered scalar fields coupled to quantum
gravitational fluctuations
\cite{frg_diffeo,Percacci-review,Percacci-Vacca1,Narain:2009qa,Narain:2009gb,Zanusso:2009bs,Vacca:2010mj}
to demonstrate that quantum gravity coupled to matter is a viable
asymptotically safe theory with a non-trivial UV-fixed point as
originally conjectured by S.~Weinberg \cite{Weinberg-2d}.

In the present paper we consider the functional RG
method in a curved space-time and focus on a quantized
scalar field $\phi$ coupled to a background with classical metric
$g_{\mu\nu}$. As we are mainly interested in the matter sector,
our attention will be mostly concentrated on the running of a
non-minimal coupling function $F(\phi)$ which directly couples to the
scalar-curvature through the interaction $F(\phi)R$. This truncation
generalizes the more familiar scalar-curvature interaction $\xi \phi^2 R$,
which was previously explored with non-perturbative methods
in \cite{AIP-EJPC}. The generalization leading
to $F(\phi)R$ and, in general, to non-polynomial self-interactions is
interesting and was previously discussed in \cite{BKK} in the framework
of effective quantum field theory. The FRG approach offers a natural framework
for dealing with such a truncation of the effective action.

It is well-known that in the conventional perturbative approach, the
renormalization of a scalar theory with $\xi \phi^2 R$ - interaction
has the following properties:
\begin{itemize}
\itemsep=0mm

\item
The presence of a term $\propto\xi \phi^2 R$ is necessary for
renormalizability of the theory. In particular, this means that the
$\be$-function for $\xi$ is non-zero, except at the fixed point. In
one-loop order of perturbation theory the fixed point value is
$\xi_*=1/6$ in four dimensions (see, e.g., \cite{book}).
This value corresponds to the local conformal symmetry of the classical
theory, and for the $d$-dimensional space the same symmetry requires
$\xi_*=(d-2)/(4d-4)$. Let us note that the conformal fixed
point is known only in four dimensions, because, for instance, in
odd-dimensional spaces the one-loop beta-functions vanish and
the results at two loops are not available. In the two-dimensional
case $\xi_*=0$ is a fixed point.

\item
In all orders of the loop expansion the $\be$-functions for the
coupling constants of the theory (such as $\la$ in the
$\la\phi^4$-interaction case) do not depend on $\xi$, while the
$\be$-function for $\xi$ is given by a polynomial expansion in
these coupling constants corresponding to the expansion in loops.
In the Minimal Subtraction scheme-based RG
the $\be$-function for $\xi$ is mass-independent. But a
dependence on the mass is seen in the momentum-subtraction
scheme of renormalization \cite{Bexi}.

\item
The renormalization of the parameters of the vacuum action
(depending on the background $g_{\mu\nu}$) depends on coupling constants
\emph{and} on $\xi$, while the inverse dependence is not seen.
\end{itemize}
In other words, in the loop expansion there is a hierarchy of the
renormalization as follows:
\begin{equation} \nonumber
\mbox{minimal terms} \ \ \to \ \
\mbox{non-minimal terms} \ \ \to \ \
\mbox{vacuum terms}.
\end{equation}
Furthermore, beyond the one-loop order and in $4$ dimensions the
$\be$-function for $\xi$ is \emph{not} proportional to the difference
$\xi-1/6$. It is certainly interesting to see whether these features
can be reproduced in a non-perturbative setting based
on the FRG. In this work we employ the Wetterich equation, which probably is the most explored
among all the currently known functional RG equations.

The paper is organized as follows. In Sect.~\ref{S2} we formulate the
FRG-equations for a scalar field theory with
non-minimal interaction function $\,F(\phi)\,$ and briefly describe
the method of calculations. Since the method is quite similar
to the one which was explained in the previous work on  $\,F(\phi)=\xi \phi^2$ \cite{AIP-EJPC},
we need not present many details here. The section ends with
the explicit form of the flow equation in any dimensions
in the local potential approximation (LPA). In section \ref{sec:fps} we study the solution of the
fixed point equations and discuss the peculiarities in different
dimensions. Thereby the main emphasis is on the fixed point equations
for the non-minimal coupling. Analytical solutions in $2$ dimensions
and numerical solutions in $d>2$ dimensions are presented
and discussed in section \ref{sec:numerics}.
In the following section \ref{sec:lpapr} the flow equations for the non-minimal
coupling function with scale-dependent wave function renormalization
$Z_k$ is derived.
(the more complex equations with scale- and field-dependent wave function renormalization $Z_k(\phi)$ are given in appendix \ref{app:wfr}).
This latter improvement includes, in particular, the anomalous dimension of the field
as the logarithmic scale derivative of $Z_k$
and goes under the name of improved LPA or LPA'.
The results with $Z_k\neq 1$ are relevant in the spontaneously
broken phase with non-vanishing expectation value of the scalar field.
Finally, in section \ref{sec:wfr} we study the perturbative RG in the
vicinity of $4$ dimensions based on the truncation with scale and
field dependent wave function renormalization. In leading order
of the $\epsilon$-expansion we calculate the critical exponents
for a scalar field coupled non-minimally to gravity. In section
\ref{S5} we draw our conclusions and discuss some perspectives
for further work on the FRG in curved space.

\section{Scalar field in curved space with non-minimal coupling}
\label{S2}

The classical action of a single real scalar field in a curved space
has the form
\begin{equation}
S=\int \sqrt{g}\,\left\{-\frac{1}{2}\phi\varDelta_g\phi
+RF(\phi)+V(\phi)\right\}
\,+\,
S^\mathrm{grav}[g]\,.
\label{act1}
\end{equation}
Here and in what follows we assume Euclidean signature for the metric $g_{\mu\nu}$, denote the covariant Laplace operator with $\Delta_g$,
and use the notation $\int \sqrt{g}=\int \D^n x\sqrt{g(x)}$.
Our purpose is to explore the quantum effects of a scalar field, while
the metric will be regarded as a classical external field. The classical
action (\ref{act1}) involves a non-minimal term, which is known to be
necessary for renormalizability in curved space. In the present paper
we are mainly interested in the non-perturbative running of the
non-minimal coupling function $F(\phi)$.

The ansatz for the scale dependent effective action is
\begin{eqnarray}
\Gamma_k&=&
\int \sqrt{g}\,\Big\{-\ha
Z_k(\phi)\,\phi\varDelta_g\phi
+RF_k(\phi)+V_k(\phi)\Big\} \nonumber \\
&&  +
\Gamma_k^\mathrm{grav}[g]\,, \label{act3}
\end{eqnarray}
and it includes a scale dependent effective potential $V_k(\phi)$,
a scale-dependent non-minimal coupling function $F_k(\phi)$
and a scale-dependent wave function renormalization $Z_k(\phi)$.
Indeed, only in section \ref{sec:wfr} and appendix \ref{app:wfr} we do allow for a field-dependent
$Z_k$, which in general has a rather lengthy flow equation. Therefore
we will derive the flow equations first in the LPA'-approximation
with scale-dependent but otherwise constant  $Z_k$.
The lengthy calculation for the case of a nonconstant wave-function renormalization
is separated into appendix \ref{app:wfr}.


Due to the above mentioned hierarchy of renormalization, we expect
that the RG flow of the non-minimal function $F_k$ does not
depend on the parameters in $\Gamma_k^\mathrm{grav}[g]$ and
can be explored separately. Of course, the purely gravitational
contribution, which is not considered in the present work,
is of relevance for the intensively studied
asymptotic safety scenario \cite{frg_diffeo,Percacci-Vacca1}.

As invariant cutoff action we choose
\begin{eqnarray}
\Delta S_k
&=&
\ha\int \sqrt{g} \,\phi R_k(-\varDelta_g)\phi \nonumber
\\ && \mtxt{with}
R_k(-\varDelta_g)=Z_k r_k(-\varDelta_g)\,.
\label{act5}
\end{eqnarray}
$R_k$ must have the well-known properties of a cutoff function
\cite{FRG-review} and will be specified later (note that differently
from the scalar curvature $R$, the cutoff-function $R_k$ is always shown with the
subscript $k$). Next we introduce the anomalous dimension
\begin{equation}
\eta_k=-\frac{k\partial_k Z_k}{Z_k}=-\frac{\partial_t Z_k}{Z_k},
\mtxt{where}
t=\log\frac{k}{\Lambda}\,.
\label{act7}
\end{equation}
The left hand side of the flow equation
\begin{equation}
\partial_t \Gamma_k[\phi]=\ha \tr \left(
\frac{\partial_t R_k}{\Gamma^{(2)}_k[\phi]+R_k}\right)\label{wetterich}
\end{equation}
is simply given by
\begin{eqnarray}
\partial_t\Gamma_k&=&\int
\sqrt{g}\,\Big\{
\frac12\,\eta_k Z_k\,\phi\varDelta_g\phi
+R\partial_t F_k(\phi)
+\partial_t V_k(\phi)\Big\} \nonumber\\
&& +\partial_t\Gamma_k^\mathrm{grav}[g].
\label{act9}
\end{eqnarray}
In the flow equation we also need the second functional derivative of
the effective action (\ref{act3}) with respect to the scalar field
\begin{equation}
\Gamma_k^{(2)}=-Z_k\varDelta_g+R F_k''(\phi) +V''_k(\phi)\,, \label{act11}
\end{equation}
and the variation of the cutoff
\begin{equation}
\partial_t R_k=Z_k\left(\partial_t r_k-\eta_k r_k\right).
\label{act13}
\end{equation}
Thus for our truncation the {\it r.h.s.} of the flow equation takes the form
\begin{eqnarray}
&& \ha \tr \left(
\frac{\partial_t R_k}{\Gamma^{(2)}_k[\phi]+R_k}\right) \label{act15} \\
&& \quad=
\ha \tr \Big\{
\frac{Z_k(\partial_t-\eta_k)r_k(-\varDelta_g)}
{-Z_k\varDelta_g+RF_k''(\phi)+V''_k(\phi)+Z_k r_k(-\varDelta_g)}\Big\}\,. \nonumber
\end{eqnarray}
To compare with the \emph{l.h.s.} of the flow equation in (\ref{act9})
we expand this expression in powers of the scalar field and curvature.
Therefore we set
\begin{eqnarray}
V_k(\phi)&=& V_{k}(0)+\frac{V''_k(0)}{2}\phi^2+W_k(\phi)\,, \nonumber \\
V''_k(\phi) &=& V_k''(0) + W''_k\,,
\label{act17}
\end{eqnarray}
where $W_k(\phi)$ contains cubic and higher powers of the field.
Then we arrive at the following form of the \emph{r.h.s.},
\begin{equation}
\ha \tr \left(
\frac{\partial_t R_k}{\Gamma^{(2)}_k[\phi]+R_k}\right)
=\ha\tr\frac{B_k(-\varDelta_g)}{P_k(-\varDelta_g)+\Sigma_k}\,,
\label{act19}
\end{equation}
where, following \cite{AIP-EJPC}, we introduced the abbreviations
\begin{align}
B_k(-\varDelta_g) & =
\pa_t r_k(-\varDelta_g) - \eta_k r_k(-\varDelta_g)\,,
\label{act21}
\\
P_k(-\varDelta_g) & =
-\varDelta_g+r_k(-\varDelta_g)+\frac{V_k''(0)}{Z_k}\,,
\label{act23}
\\
\Sigma_k(\phi,R)&=\frac{1}{Z_k}
\big[RF''_k(\phi)+W''_k(\phi)\big]\,.
\label{act25}
\end{align}
We will expand the \emph{r.h.s.} on (\ref{act19}) in a power series
in $\Sigma_k$ and thereby use $[B_k,P_k]=0$.
However, for a inhomogeneous field and curvature the
spacetime-dependent $\Sigma_k$ does not commute with $B_k$ and
$P_k$. But we still can write down the Neumann series
\begin{eqnarray}
\tr\Big\{\frac{B_k}{P_k+\Sigma_k}\Big\}
&=&
\sum_{m\geq 0}(-1)^m\, \tr \Big\{
Q_{k,1}\left(P_k^{-1}\,\Sigma_k \right)^m\Big\}\,, \nonumber
\\ 
Q_{k,m}&=&\frac{B_k}{P_k^m}.
\label{act29}
\end{eqnarray}
To simplify the notations we skip the argument $-\varDelta_g$ of $B_k,P_k$ and
$Q_{k,m}$ as well as the arguments $\phi$ and $R$ of $\Sigma_k$.

The traces appearing in (\ref{act29}) can be computed via the heat
kernel of the covariant Laplacian. The details of
this procedure were described in \cite{off-diagonal,AIP-EJPC} and we just
present the result for the optimized regulator function
$r_k(s)=(k^2-s)\theta(k^2-s)$ \cite{Litim:2000ci}.
The expression for the trace in Eq.~(\ref{act29}) is
\begin{eqnarray}
Q_{k,m}(s)&=&\frac{2k^2-(k^2-s)\eta_k}{\Delta_k^m}
\,\theta(k^2-s),\nonumber
\\ 
\Delta_k&=&k^2+\frac{m_k^2}{Z_k}\,.
\label{act31}
\end{eqnarray}
Let us now consider the asymptotic small-$t$ expansion of
$\exp(t\varDelta_g)$,
\beq
\E^{t\varDelta_g}
=\frac{1}{(4\pi t)^{d/2}}\left(A_0+tA_1+t^2A_2+\dots\right),
\label{act33}
\eeq
where the Schwinger-DeWitt coefficients have the well-known form,
\begin{eqnarray}
&& A_0=1,\quad
A_1=\frac{1}{6}R,
\quad \label{act35}\\
&& A_2=\frac{1}{180}\Big(R_{\mu\nu\alpha\beta}R^{\mu\nu\alpha\beta} \nonumber \\
&& \qquad\qquad\qquad
-R_{\mu\nu}R^{\mu\nu}+6\varDelta_g R+\frac{5}{2}R^2\Big).
\end{eqnarray}

With the help of the asymptotic expansion (\ref{act33}) the operators
defined $Q_{k,m}$ in (\ref{act29}) (the reader may consult \cite{AIP-EJPC} for
more details), one arrives at the series expansions in position space,
\begin{eqnarray}
&& \langle x\vert Q_{k,m}(-\varDelta_g)\vert x\rangle=
\frac{2}{(4\pi)^{d/2}}\frac{1}{\Delta_k^m}  \label{act37}\\
&&\qquad \times\sum_{n\geq0} \frac{k^{d-2n+2}}{\Gamma(d/2-n+1)}
\left(1-\frac{\eta_k}{d-2n+2}\right)\,A_n(x)\,,\nonumber
\end{eqnarray}
where $\Delta_k$ has been introduced in (\ref{act31}). Note that for even
$d$ the series terminate since $1/\Gamma$ has zeros on the set of
non-positive integers. At the same time our linear in curvature
approximation is such that only the terms $\,n=0,1\,$ are relevant.

\subsection{Local potential approximation}
\label{subsec:lpa}

In a first step we consider the local potential approximation (LPA)
with constant $\phi$ and constant scalar curvature $R$. Later we shall see how
space-time dependent fields may modify the results. In the LPA no
terms with derivatives of the field $\phi$ appear in the \emph{r.h.s.}
of the flow equation and hence $\partial_t Z_k$ vanishes in this
approximation. The generalization to theories with
Spontaneous Symmetry Breaking and non-trivial wave-function
renormalization will be dealt with later on.

In the approximation considered $P_k$ commutes with $\Sigma_k$
and the Neumann series (\ref{act29}) simplify to
\begin{eqnarray}
\tr\left(\frac{B_k}{P_k+\Sigma_k}\right)
&=&\sum_{m=0}^\infty (-1)^m\,\tr Q_{k,1}P_k^{-m}\Sigma_k^m \nonumber \\
&=&\sum_{m=0}^\infty (-1)^m\,\tr Q_{k,m+1}\Sigma_k^m.\label{lpa1}
\end{eqnarray}
Inserting the expansion (\ref{act37}) with $\eta_k=0$ for the
operators $Q_{k,m+1}$, one obtains a double sum over $m$
and $n$. The sum over $m$ can be carried out and provides as
intermediate result
\begin{eqnarray}
&& \ha\tr\left(\frac{B_k}{P_k+\Sigma_k}\right)  \label{lpa3}\\
&&\qquad  =\frac{1}{(4\pi)^{d/2}}\sum_{n\geq 0}\frac{k^{d-2n+2}}{\Ga(d/2-n+1)}
\tr \left(\frac{A_n}{\Delta_k+\Sigma_k}\right).
\nonumber
\end{eqnarray}
In the given truncation only the terms with $n=0$ and $1$ contribute, such
that the relevant part of the \emph{r.h.s.} of the flow equation
is
\begin{eqnarray}
&&\ha\tr\left(\frac{B_k}{P_k+\Sigma_k}\right) \nonumber\\
&&\quad = \mu_d \mathrm{Vol}\Big(
\frac{k^{d+2}}{\Delta_k+\Sigma_k}
+\frac{d}{12}\frac{k^d}{\Delta_k+\Sigma_k}\,R\Big)
+\dots\,,\label{lpa5}
\end{eqnarray}
where Vol denotes the space-time volume.
In addition we introduced the geometric factor
\bea
&& \mu_d=\frac{1}{(4\pi)^{d/2}\Gamma(\frac{d}{2}+1)},
\nonumber \\
&& \mbox{e.g.,} \quad \mu_2=\frac{1}{4\pi}
,\quad
\mu_3=\frac{1}{6\pi^2}
,\quad
\mu_4=\frac{1}{32\pi^2}\,.
\label{lpa7}
\eea
Expanding $(\Delta_k+\Sigma_k)^{-1}$ in Eq.~\eqref{lpa5} in powers
of of the Ricci scalar, only the two leading terms contribute in our truncation,
and we get
\bea
&& \ha\tr\left(\frac{B_k}{P_k+\Sigma_k}\right)
= k^d \mu_d \mathrm{Vol}
\Bigg[k^2\,\Big(\frac{1}{k^2+V''_k} \nonumber \\
&&\qquad\qquad -\frac{R F''_k}{(k^2+V''_k)^2}
\Big)+\frac{d}{12}
\frac{R}{k^2+V''_k}\,\Bigg]+\dots\,.
\label{lpa9}
\eea
Comparing with the \emph{l.h.s.} of the Wetterich equation in
(\ref{act9}) yields
\begin{align}
\partial_t V_k
&=
\mu_d k^d\frac{k^2}{k^2+V''_k}\,,
\label{lpa11a}
\\
\partial_t F_k
&=
\mu_d k^d\left(\frac{d}{12}
\frac{1}{k^2+V''_k}-\frac{k^2 F''_k}{(k^2+V''_k)^2}\right)\,.
\label{lpa11b}
\end{align}
The first of these equations is exactly the same as in flat
space-time, while the second one has no analogs in the flat-space limit.
It is easy to see that these two flow equations
imply that for even functions $V_\Lambda(\phi)$
and $F_\Lambda(\phi)$ at the UV-cutoff the scale dependent
functions $V_k$ and $F_k$ stay even at all scales $k$.

Note that the flow of the effective potential $V_k(\phi)$
is independent of the non-minimal coupling function $F_k(\phi)$
and is exactly the same as in flat space-time. However it determines
the running of the non-minimal coupling function $F_k(\phi)$.

\section{Fixed point solutions}
\label{sec:fps}
To localize the fixed points of the RG flow we introduce
the dimensionless field $\chi$, potential $v_k(\chi)$ and coupling function $f_k(\chi)$:
\bea
&& \phi=k^{(d-2)/2}\,\chi,\qquad
V_k(\phi)=k^d\, v_k(\chi)\,, \nonumber \\
&& \qquad\qquad F_k(\phi)=k^{d-2}f_k(\chi).
\label{gen7}
\eea
As a rule we denote dimensionful parameters and potentials
by capital letters and the corresponding dimensionless quantities by
small letters. The only exception is $\phi$ and $\chi$.
By means of the identities
\bea
\partial_\phi^2&=&k^{2-d}\partial_\chi^2\,, \nn \\
\partial_t V_k
&=& k^d\left(\partial_t v_k
+ d v_k
- \frac{d-2}{2} \,\chi\,\partial_\chi v_k\right)\,,
\label{afp3}
\eea
and similarly for $\partial_t F_k$,
the flow equations for the dimensionless quantities take the form
\bea
 && \partial_t v_k
+d v_k-\frac{d-2}{2}\chi v_k'
= \frac{\mu_d}{1+v''_k}\,,\label{afp5a}
\\
&&\partial_t f_k+(d-2)f_k -\frac{d-2}{2}\,\chi f_k'\nn \\
&&\qquad\qquad\qquad\qquad
=\frac{d}{12}\frac{\mu_d}{1+v''_k}
-\frac{\mu_d f''_k}{(1+v''_k)^2}\,.
\label{afp5b}
\eea
Scaling solutions for the effective potential and the non-minimal
coupling function are $k$-independent
\emph{global} solutions $v_*$ and $f_*$ of (\ref{afp5a}) which
generalize the notion of a RG fixed point.
We shall denote these fixed point solutions by a star in the following.
The fixed point equations are
\begin{align}
v''_*&=\frac{2\mu_d}{2dv_*-(d-2)\chi v'_*}-1\,,  \label{afp10}
\\
f''_*&=\frac{d}{12}(1+v_*'')+\frac{d-2}{2\mu_d}
(1+v_*'')^2\left(\chi f_*'-2f_*\right).
\label{afp11}
\end{align}
If the $v_k$ and $f_k$ are even functions at the cutoff, then
they are even at any scale. Thus we assume the expansions
\bea
f_*(\chi)&=&f_*(0)+\frac{\xi_*}{2} \chi^2+\dots\,,
\nn \\
v_*(\chi)&=&v_*(0)+\frac{m_*^2}{2}\chi^2+\dots\,.\label{fin1}
\eea
The constant term $f_*(0)$ relates to the dimensionless gravitational
constant which feeds into the purely gravitational sector, which is
not considered in the present work. Later we shall see that the fixed
point value $\xi_*$ depends on this constant.

Inserting these expansions into the flow equation (\ref{afp5b})
(not the equation (\ref{afp11}) where we solved for $f_*''$)
with $\partial_t v_*=\partial_t f_*=0$ and comparing coefficients
of $\chi^0$ relates $\xi_*\equiv f_*''(0)$ to $m_*$ and $f_*(0)$,
\begin{equation}
\xi_*=\frac{d}{12}(1+m_*^2)-\frac{d-2}{\mu_d}(1+m_*^2)^2\,f_*(0)\,.\label{fin3}
\end{equation}
Comparing coefficients of $\ch^2$
relates $\xi_*$ to the fourth derivatives of $v_*$ and $f_*$,
\begin{equation}
\left(\xi_*-\frac{d}{24}(1+m_*^2)\right)v_*''''(0)=\frac{1+m_*^2}{2}
f_*''''(0)\,.\label{fin5}
\end{equation}
For even functions $v_*,f_*$ the fixed point equation (\ref{afp11})
implies that $f''''_*(0)$ is proportional to $v''''_*(0)$. Using this
relation in (\ref{fin5}) gives rise to the simpler
result (\ref{fin3}).

\subsection{Gaussian fixed points}
In all dimensions there exist Gaussian fixed point solutions
of (\ref{afp10}) and (\ref{afp11}) with constant scaling potential $v_*$.
Then $m_*$ and $v_*''''(0)$ both vanish and
(\ref{fin5}) does not yield information about $\xi_*$,
but instead implies $f_*''''(0)=0$. We conclude that at a
\emph{Gaussian fixed point} the non minimal function $f_*$ is a
polynomial of degree $2$. The relation (\ref{afp10}) determines
the constant $v_*$ and (\ref{afp11}) determines the coefficients
of the quadratic polynomial $f_*$:
\bea
 && v_*(\chi)=\frac{\mu_d}{d},\qquad
  f_*(\chi)=\frac{\xi_*}{2}\chi^2+f_*(0), \nn \\
 &&\qquad \qquad \xi_*=\frac{d}{12}-\frac{d-2}{\mu_d}f_*(0)\,.
 \label{afp13}
\eea
Only in two dimensions is the non-minimal parameter
independent of the normalization $f_*(0)$. In higher dimensions
$f_*(0)$ shifts the value of $\xi_*$

\subsection{Non-Gaussian fixed points}

Let us assume that there exists an \emph{interacting (non-Gaussian) fixed point}
with non-vanishing self-coupling $v_*''''(0)$ and truncate the non-minimal function $f_*$ to an even polynomial of degree $2$
as in (\ref{afp13}). Then $f_*''''(0)=0$ and (\ref{fin5})
determines $\xi_*$ which we insert into (\ref{fin3}) to find $f_*(0)$:
\bea
&& \xi_*=\frac{d}{24}(1+m_*^2),  \label{afp12b}\\
&& f_*(0)=\frac{d\mu_d}{24(d-2)(1+m_*^2)}\,, \mtxt{if}
f_*''''(0)=0,\;\; d>2\,. \nn
\eea
Two dimensions are special since (\ref{fin3}) yield for any
fixed point -- interacting or non-interacting -- the simple relation
\begin{equation}
\xi_*=\frac{1}{6}(1+m_*^2),\qquad d=2,\label{fin7}
\end{equation}
independent of $f_*(0)$.
Then (\ref{fin5}) implies that at a non-Gaussian fixed point
with non-vanishing $v_*''''(0)$ the non-minimal function
$f_*$ can not possibly be a polynomial of degree $2$.

In dimension higher than two a truncation with quadratic $f_*$
is inconsistent if $f_*(0)=0$. In other words
$f_*(0)$ and $f_*''''(0)$ cannot both vanish
at a non-trivial fixed point. The truncation to a
quadratic polynomial $f_*(\chi)$  has been discussed in detail for $d=4$
in \cite{AIP-EJPC}. Let us now discuss the non-trivial fixed points which
are expected to exist in $2\leq d<4$ dimensions in more detail.
Actually for dimensions $3\leq d<4$ there exists one fixed point
and below $3$ dimensions we expect a proliferation of fixed
points with decreasing $d$.
\paragraph{4 dimensions:}
If there would exist an interacting fixed point in $4$ dimensions
(which probably is not the case) then we have for the
\emph{truncation} $f_*=f_*(0)+\xi_*\chi^2/2$
according to (\ref{afp12b})
\begin{equation}
\xi_*=\frac{1}{6}(1+m^2_*),
\label{afp17}
\end{equation}
which may deviate from the classical value $1/6$. This should be compared
with the prediction of the standard Minimal Subtraction scheme-based
one-loop RG for $\xi$,  where a mass-dependence is not seen.
Let us note that the mass-dependent $\be$-functions are encountered in
the physical (e.g., momentum-subtraction) renormalization schemes,
including the non-minimal parameter $\xi$.
In principle, starting from 3 loops the beta-function for $\xi$ is not
proportional to $\xi-1/6$, as is known from \cite{Hathrell-82}.

\paragraph{3 dimensions:}
The fixed point equation (\ref{afp10}) takes the form
\begin{equation}
v''_*=
\frac{1}{3\pi^2}\frac{1}{6v_*-\chi v_*'}-1\,,\label{afp27}
\end{equation}
and admits a nontrivial scaling solution. Indeed, a numerical
study reveals that only for the initial condition
$v_*''(0)=m_*^2=-0.18605$ a non-trivial \emph{global
solution} of the (singular) differential equation exists \cite{Litim2,Wipf}.
As a result the critical $\xi_*$ in (\ref{afp12b}) is slightly
smaller than the classical value $1/8$ (corresponding to the
conformal coupling of a scalar field to gravity),
\begin{equation}
\xi_*=\frac{1}{8}(1+ m^2_{*})=
0.10174<\xi_\mathrm{classical}=0.125\,.\label{afp31}
\end{equation}

\paragraph{From $3$ to $4$ dimensions:}
In four dimensions there is probably only the Gaussian fixed
point solution for a scalar field \cite{wilsonfroehlich}. In LPA it has
constant potential
\begin{equation}
v_*\,=\frac{1}{128\pi^2}
\mtxt{and}
\xi_*=\frac{1}{3}-64\pi^2 f_*(0)\,.\label{afp26}
\end{equation}
Let us see what happens when we approach the upper critical
dimension $d=4$ from below. Since the FRG can be formulated in
arbitrary space-time dimensions, we may continuously increase
$d$ from $3$ to $4$ and study the limit of $\xi_*$
when $d\to 4$. In all dimensions $3\leq d<4$ there exists
\emph{one non-trivial fixed point solution} $v_*(\chi)$ with non-vanishing
$v_*''''(0)$ and $m_*^2<0$.
Following \cite{codello2,Hellwig:2015woa} we numerically solved the singular fixed point equation
(\ref{afp10}) for the dimensions given in Table~\ref{xistar}
with the shooting method. An even solution depends only on its initial
value $v_*(0)$ or equivalently on its initial curvature
$v_*''(0)=m_*^2$. For a wrong initial condition $m_*^2$
the solution of the singular differential equation
(\ref{afp10}) becomes singular at a finite value $\chi_\mathrm{max}$ of the
field.
We fine-tuned the mass-parameter $m_*^2$ such that the solution
extends to a maximal value of $\chi_\mathrm{max}$. If one continues with
this fine-tuning then $\chi_\mathrm{max}$ finally increases until one
(in principle) obtains a globally well-defined potential.
After the global solutions $v_*$ is known one can proceed with
solving the regular differential equation (\ref{afp11}).
The approximate fixed point values of $m_*^2$ and $\xi_*$ are listed
in Table~\ref{xistar}.
\begin{table*}
\begin{tabular}{lrrrrrrr}\hline
 d&3&3.1&3.2&3.3&3.4&3.5&3.6\\
$m_*^2$&-0.18605&-0.15662&-0.13002&-0.10609&-0.08462&-0.06544&-0.04843\\
$\xi_*$&0.10174&0.10894&0.11600&0.12291&0.12968&0.13629&0.14274\\  \hline
 d&3.7&3.8&3.9&3.95&3.98&3.99&3.999\\
$m_*^2$&-0.033457  &-0.020430
&-0.009283&-0.004406&-0.0017053&-0.0008430&-0.00008343\\
$\xi_*$&0.149009&0.155099&0.160992&0.163858&0.165551&0.166110&0.166611
\\ \hline
\end{tabular}
\caption{The curvature of the scaling potential with corresponding
non-minimal parameter $\xi_*$ in various dimensions between $3$ and $4$.
The results are plotted in Fig. \ref{figuremxi}.}
\label{xistar}
\end{table*}

\begin{figure}[h]
\centering
\includegraphics[width=8cm]{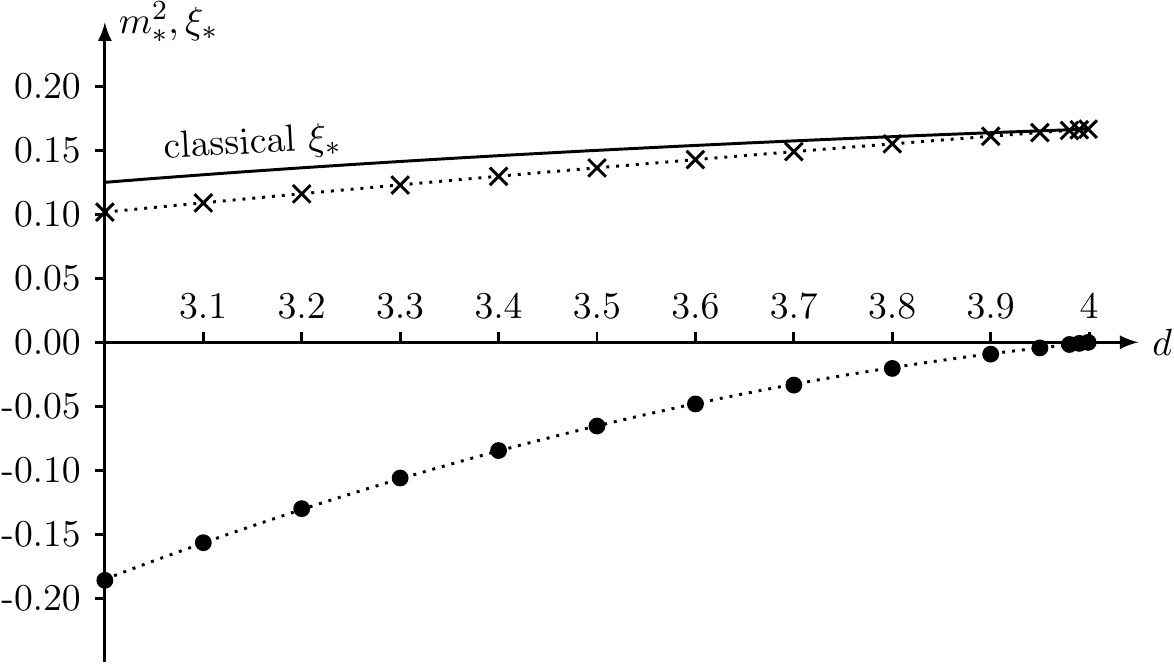}
\caption{The numerically determined values $m_*^2=v_*''(0)$
(marked by $\bullet$) with corresponding
non-minimal parameters $\xi_*$ (marked by $\times$)
in dimensions ranging from $3$ to $4$.
The solid curve shows the classical conformal coupling.
The dotted curves are fits by polynomials of degree $2$.}
\label{figuremxi}
\end{figure}

For calculating the non-minimal coupling $\xi_*$ we used the relation
(\ref{afp12b}) which applies to interacting fixed points in a
truncation with quadratic $f_*$.
\noindent
These values are depicted in Figure~\ref{figuremxi}, where we also
plotted the classical conformal couplings and the interpolating
polynomials of degree two for the calculated mass parameters and
non-minimal parameters, given by
\begin{align}
\begin{split}
\xi_*(d)&=-0.0084105\,d^2+0.123892\,d-0.194295\,, \\
m_*^2(d)&=\hskip4mm -0.10973\,d^2+0.95288\,d\;-\;2.05629\,.
\end{split}\label{afp32}
\end{align}
In $4$ dimensions the interpolating polynomials yield the values $\xi_*=1.667$ and $m_*=-4\cdot 10^{-4}$.
The value for $\xi_*$ agrees with the prediction (\ref{afp17}) for an
interacting fixed point with $m_*=0$ or with the prediction (\ref{afp26})
for a Gaussian fixed point with $f^{-1}_*(0)=384\pi^2$.

From the calculated values at $d=4-\epsilon$ with small $\epsilon\leq 0.3$
we extracted via an interpolation by a polynomial
of degree $2$ the following $\epsilon$-expansions:
\bea
\xi_*(4-\epsilon)&\approx&
0.166669-0.055753\,\epsilon-0.010395\,\epsilon^2+\dots \nn \\
&\approx&\;
\frac{1}{6}-\frac{\epsilon}{18}-\frac{\epsilon^2}{96}+\dots\,,\nn \\
m^2_*(4-\epsilon)&\approx& 3.2497\cdot 10^{-7}-0.083448\,\epsilon-0.093580\,\epsilon^2
+\dots \nn \\
& \approx& -\frac{\epsilon}{12}-\frac{468\epsilon^2}{5000}\,.
\label{afp34}
\eea
The $\epsilon$-expansion with scale and field-dependent wave function
renormalization is reconsidered in section \ref{sec:wfr}.

\section{Exact and numerical solutions}\label{sec:numerics}
Using the fixed point equation (\ref{afp10}) for $v_*$, the
differential equation (\ref{afp11}) for $f_*$ can be written as
\begin{equation}
f_*''=
\frac{2\mu_d}{2d v_*-(d-2)\chi v_*'}
\left[\frac{d}{12}+\frac{(d-2)(\chi f'_*- 2 f_*}{2dv_*-(d-2)\chi v_*'}\right].
\label{gen11}
\end{equation}
This form is convenient for numerical studies.

\subsection{Analytic solutions in $2$ dimensions}
In two dimensions there exist an infinite set of non-perturbative
fixed points solutions of the fixed point equation \cite{morris}
\begin{equation}
v''_*=\frac{1}{8\pi}\frac{1}{v_*}-1\,.
\label{afp33}
\end{equation}
Multiplying with $v_*'$ we immediately find a first
integral. For an even scaling potential
it reads
\begin{equation}
\frac{1}{2}v_*'^{\,2}(\chi)=\frac{1}{8\pi}\log\frac{v_*(\chi)}
{v_*(0)}-\left(v_*(\chi)-v_*(0)\right)
\,.\label{afp35}
\end{equation}
For a real potential the left hand side is non-negative
which implies
\begin{equation}
\log\frac{v_*(\chi)}{v_*(0)}
-8\pi\left(v_*(\chi)-v_*(0)\right)\geq 0\,.
\label{apf39}
\end{equation}
By inspection one sees that for a positive initial value $v_*(0)$
the left hand side vanishes at two positive values
$v_*(\chi)$ and is positive between these two nodes only. This
means that for a positive $v_*(0)$ the potential $v_*(\chi)$
is bounded from below and from above.
There are two possibilities \cite{morris},
\begin{align}
m_*^2>0&\Longrightarrow v_{*}(0)
\leq v_*(\chi)
\leq v_{*\mathrm{max}}, \nonumber \\
-1<m_*^2<0&\Longrightarrow v_{*\mathrm{min}}\leq
v_*(\chi)\leq v_*(0)\,.
\label{afp41}
\end{align}
The fixed point equation (\ref{afp33}) relates the potential
and its curvature at the origin,
\begin{equation}
 v_*(0)=\frac{1}{8\pi}\frac{1}{1+m_*^2}\,,
 \label{afp43}
\end{equation}
such that $v_*(0)$ varies between $0$ and $1/8\pi$
for the first class of solutions in (\ref{afp41})
and is bigger than $1/8\pi$ for the second
class. These bounded solutions show an oscillatory behavior.
On the other hand, for a \emph{negative} $v_*(0)$ the left
hand side of (\ref{apf39}) has only one node and $v_*(\chi)$
is negative for all $\chi$ and unbounded from below.
This unstable solutions will be discarded on physical grounds.

\begin{figure}[h]
\centering
\includegraphics[width=8cm]{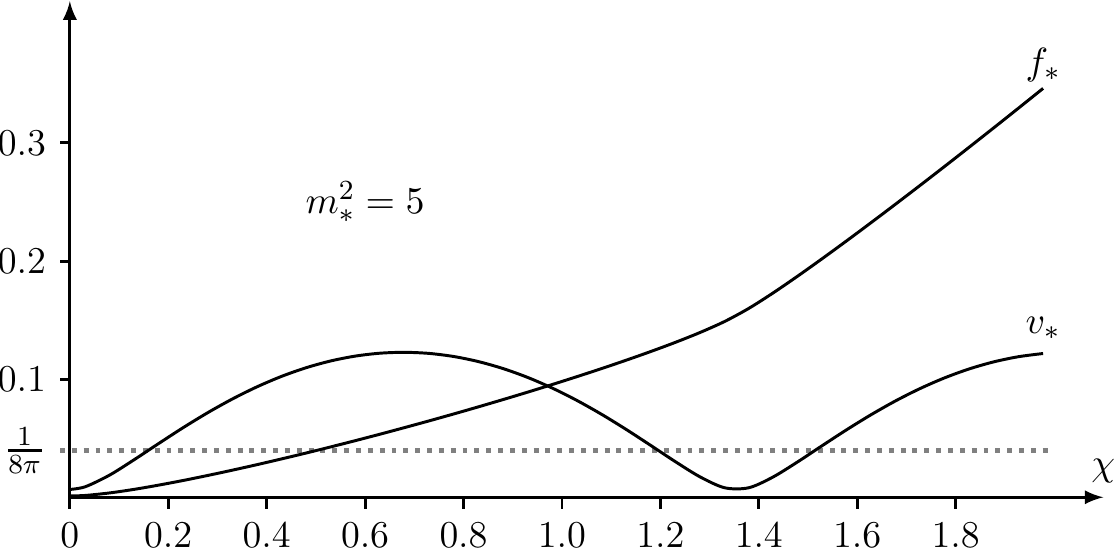}\\[3mm]
\includegraphics[width=8cm]{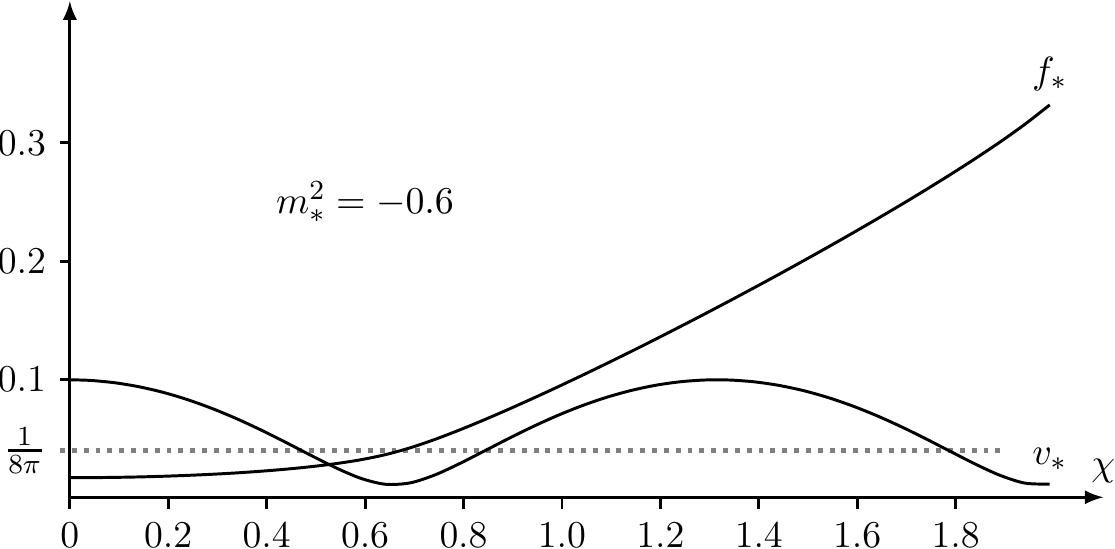}
\caption{Two periodic scaling potentials $v_*(\chi)$ in two
dimensions (without wave function renormalization) with corresponding
coupling function $ f_*$. We set $f_*(0)=0$ in (\ref{gen19}).}
\label{figure2d}
\end{figure}

The inverse function $\chi=\chi(v_*)$ of a solution $v_*(\chi)$ of
 (\ref{afp33}) is given
by the integral \cite{codello2}
\begin{equation}
 \chi(v_*)=\sqrt{4\pi}\int_{v_*(0)}^{v_*}
 \frac{\D u}{\sqrt{\log u/v_*(0)-
 8\pi(u-v_*(0))}}\,.\label{afp45}
\end{equation}
In $2$ dimensions the fixed point equation (\ref{afp11}) for $ f_*$
simplifies considerably and can
easily be solved in terms of the scaling potential.
Even solutions have the form
\begin{equation}
 f_*(\chi)=\frac{\chi^2}{12}+\frac{v_*(\chi)}{6}+
f_*(0), \label{gen19}
\end{equation}
with scaling potential given in (\ref{afp45}).
Each fixed point comes with its own periodic scaling potential
$v_*$, non-minimal coupling function $f_*$ and \emph{positive}
non-minimal parameter,
\begin{equation} \xi_*\equiv
f_*''(0)=\frac{1}{6}(1+m^2_*)=\frac{1}{48\pi\, v_*(0)}\in [0,\infty]\,.\label{afp47}
\end{equation}
The \emph{classical} conformal coupling in two dimensions is $\xi_*=0$
and it is attained for $m_*^2=-1$. According to
(\ref{afp41}) this is the value for which
the periodic scaling solutions with $m_*^2>-1$ cease to
exist and one encounters the unstable scaling solutions with $m_*^2<-1$.
Two typical solutions of the flow equation
(\ref{afp33}) with associated $f_*$ in (\ref{gen19})
are depicted in Fig. \ref{figure2d}.
We normalized $f_*$ by setting $f_*(0)=0$.
The potential $v_*$ in the upper figure
has positive curvature $m_*^ 2$ at the origin and the one in the lower
figure has negative $m_*^2$. Note that the asymptotic form of $ f_*$
for large values of the field is independent of the scaling potential.\footnote{This corresponds to the fact that the $d=2$ theory in a usual perturbative approach is renormalizable for any functions $v$ and $f$.}

Note that in $2$ dimensions we did not truncate $f_*(\chi)$ to a
quadratic polynomial. Without wave function renormalization we
observe a continuum
of scaling potentials in two dimensions
and correspondingly any value of $\xi_*$
between $0$ and $\infty$ seems to be possible.
But we expect sizable corrections to the
fixed point solutions if we include a wave function renormalization.
With wave function renormalization
one finds a discrete set of scaling potentials and
correspondingly a discrete series of fixed point values $\xi_*$,
see section \ref{sec:lpapr}.
\subsection{Three and four dimensions}
\begin{figure}[h]
\includegraphics[width=8cm]{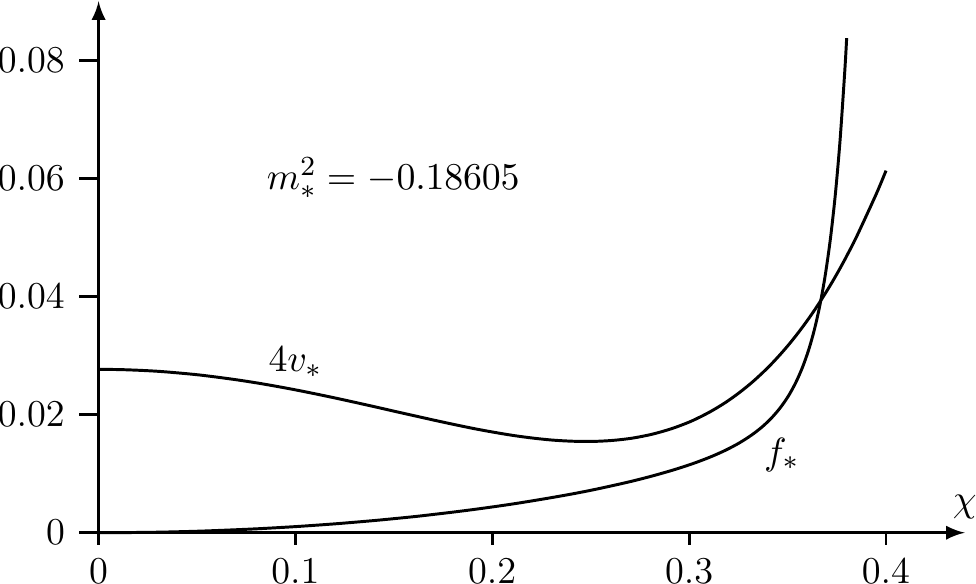}
\caption{The scaling potential $v_*(\chi)$ in three dimensions
(without wave function renormalization) and the
corresponding coupling function $ f_*$ with initial condition
$ f_*(0)=0$.}
\label{figure3d}
\end{figure}

As discussed earlier, in \emph{three dimensions} there is one
non-trivial even scaling potential $v_*$ characterized
by its (fine-tuned) mass-parameter.
First we solved (\ref{afp27}) for the scaling potential and in a
second step obtained the fixed point coupling function $ f_*$ from
(\ref{gen11}), which in $3$ dimensions reads
\begin{equation}
f''_*=\frac{1}{3\pi^2}\,\frac{1}{6v_*-\chi v'_*}
 \left(\frac{1}{4}
 -\frac{2 f_*-\chi f'_*}{6v_*-\chi v'_*}\right).\label{gen21}
\end{equation}
The numerical (even) solutions of the coupled system of
differential equations (\ref{afp27},\ref{gen21}) with initial
condition $ f_*(0)=0$
are depicted in Fig. \ref{figure3d}.
According to (\ref{fin3}) the non-minimal parameter $\xi_*$
depends on the unspecified fixed point value of the
dimensionless gravitational constant $ f_*(0)$
\begin{equation}
\xi_*=\frac{1}{4}\left(1+m_*^2\right)
\left( 1-24\pi^2 \big(1+m_*^2\big)f_*(0)\right)\,, \label{gen23}
\end{equation}
and thus is not determined by $v_*$ alone.

In \emph{four dimensions} the fixed point equation for $v_*$
admits only a constant solution belonging to the
Gaussian fixed point. The corresponding even solution in
(\ref{afp13}) contains one free parameter,
\begin{equation}
  f_*=\frac{\chi^2}{6}+ f_*(0)\left(1-32\pi^2\chi^2\right)\,.
 \label{gen25}
\end{equation}
Thus at the Gaussian fixed point the solution $f_*$
is a sum of a constant and the quadratic term. The constant term $f_*(0)$
belongs to an induced Einstein-Hilbert term, and the quadratic term is
required to provide perturbative renormalizability of the theory in curved space-time.

\section{Including the wave function renormalization $Z_k$}
\label{sec:lpapr}

We have seen  that there is no wave function renormalization in the
truncation (\ref{act3}) if an even potential $V_k$ is expanded in powers
of the field. But expanding in powers of the field may be inappropriate.
For example, the potential at intermediate scales and the
scaling solution need not be convex, contrary to the full effective
potential $V_{k\to 0}$. Indeed, in most cases the scaling solution
$v_*$ is non-convex. Since the flow is driven by fluctuations about
the minimum $\phi_{0,k}$ of the effective potential it should be
advantageous to expand both sides of the flow equation in powers of
$\delta\phi_k=\phi-\phi_{0,k}$ rather than in powers of $\phi$.
Since for an even classical potential one finds odd
powers of $\delta\phi_k$ one should allow for odd powers
in the flow equation. Then we expect a wave function renormalization
already in the truncation (\ref{act3}). Actually it is well-known that
without considering the anomalous dimension $\eta_*$ at the fixed
point one misses interesting scaling solutions in two-dimensional
systems in flat space-time \cite{delamotte}.

For systems with running $Z_k$ (and therefore non-vanishing anomalous
dimension) one easily reinstalls the $Z_k$ and $\eta_k$
dependence in the right hand side of the flow equations.
Looking at the last expression in (\ref{act15}) we see that $V_k$
and $F_k$ must be divided by $Z_k$. Going back
to (\ref{act37}) we conclude that the $n$'th term in (\ref{lpa3})
is multiplied with the factor $1-\eta_k/(d-2n+2)$. This way one
obtains the flow equations with (field independent) wave function
renormalization,
\bea
\partial_t V_k&=&\mu_d k^d\frac{k^2}{k^2+V''_k/Z_k}
\left(1-\frac{\eta_k}{d+2}\right)\,, \nn
\\
\partial_t F_k&=&\mu_d k^d\Bigg(
\frac{d}{12}
\frac{1}{k^2+V''_k/Z_k}\left(1-\frac{\eta_k}{d}\right) \nn \\
&&-\frac{k^2 F''_k/Z_k}{( k^2+V''_k/Z_k)^2}\left(1-\frac{\eta_k}{d+2}\right)
\Bigg)\,.
\label{blpa1}
\eea
For $Z_k=1$ and $\eta_k=0$ they simplify to the previously considered flow equation
(\ref{lpa11a}) and (\ref{lpa11b}).

\subsection{Scaling solutions}
To study the fixed-point solutions we introduce the dimensionless
``renormalized'' field $\chi$ and functions $v_k,f_k$ according to
\bea
&& \chi=k^{(2-d)/2} Z_k^{1/2}\,\phi,\qquad
 v_k(\chi)=k^{-d}\, V_k(\phi) \nn \\
 && \qquad f_k(\chi)=k^{2-d} F_k(\phi)\,.
 \label{blpa3}
\eea
The flow equations for the dimensionless quantities take the form
\bea
 && \partial_t v_k+dv_k-\frac{d-2+\eta_k}{2}\,\chi v_k' =\frac{\mu_d}{1+v''_k}
\left(1-\frac{\eta_k}{d+2}\right)\,, \nn \\
\label{blpa7a}
\\
&&\partial_tf_k+(d-2)f_k-\frac{d-2+\eta_k}{2}\,\chi f_k' \nn\\
&& =\frac{d}{12}
\frac{\mu_d}{1+v''_k}\left(1-\frac{\eta_k}{d}\right)
-\frac{\mu_df_k''}{(1+v''_k)^2}\left(1-\frac{\eta_k}{d+2}\right).\,
\label{blpa7b}
\eea
For $\eta_*=0$ one recovers the flow equations (\ref{afp5a}) and (\ref{afp5b})
in the LPA.
Compared to the flow equations without anomalous dimension
the ``effective space-time dimension''
appearing in the geometric terms on the left hand side is
$d+\eta_k$ instead of $d$.
Note that it is the anomalous dimension $\eta_k$ -- not the wave function
renormalization $Z_k$ -- that enters here as free parameter.
It will be determined in a later
stage when we find an algebraic equation which includes $\eta_k$.

The flow equations give rise to the fixed point equations
within the LPA':
\bea
&& dv_*-\frac{d-2+\eta_*}{2}\chi v'_*
=\frac{\mu_d}{1+v''_*}\left(1-\frac{\eta_*}{d+2}\right)\,,\label{blpa13}\\
&& (d-2)f_*-\frac{d-2+\eta_*}{2}\,\chi f_*' \nn \\
&&=\frac{d}{12}
\frac{\mu_d}{1+v''_*}\left(1-\frac{\eta_*}{d}\right)
-\frac{\mu_df_*''}{(1+v''_*)^2}\left(1-\frac{\eta_*}{d+2}\right).\,
\label{wfn1}
\eea
The flow and fixed point equation for the potential in flat space has been studied
extensively in the literature (see for example \cite{delamotte,codello2}) and hence we
focus especially on the non-minimal coupling to gravity.

In passing we note that in $2$ dimension the differential equation (\ref{wfn1})
turns into a first order equation for $f_*'$ which can be integrated.
The solution with $f_*'(0)=0$ has the form
\begin{align}
f'_*(x)&=\frac{1}{3}\,\frac{2-\eta_*}{4-\eta_*}\; e^{Q(x)}\,\int_0^x dy\,e^{-Q(y)}
\big(1+v''_*(y)\big),\nonumber\\
Q(x)&=\frac{8\pi\eta_*}{4-\eta_*}\int_0^x dy\,y\big(1+v_*''(y)\big)^2\,.
\label{wfn7}
\end{align}
In the symmetric phase the non-minimal coupling is
$\xi_*^{(s)}=f''_*(0)$. In the broken phase, where the field fluctuates about
the minimum of the effective potential, it is more reasonable
to characterize the coupling of the quantum field to
space-time curvature by $f''_k($minimum of $v_k)$.
Thus below we consider the quantities
\begin{equation}
\xi_*^{(s)}=f''_*(0)\mtxt{and}
\xi_*^{(b)}=f_*''(\chi_{0*})\,,\label{fleq8}
\end{equation}
defined at the origin and the minimum $\chi_{0*}$ of the fixed point potential $v_*$.
Both are given by a generalization of (\ref{fin3})
\bea
\xi_*^{(s,b)}&=&\Bigg(\frac{d}{12}\left(1+v_*''^{\,(s,b)}\right)
\left(1-\frac{\eta_*}{d}\right) \label{fleq9} \\
&&-\frac{d-2}{\mu_d}\left(1+v_*''^{\,(s,b)}\right)^2f_{*}^{(s,b)}\Bigg)
\left(1-\frac{\eta_*}{d+2}\right)^{-1}\,,\nn
\eea
where $v_*''^{\,(s,b)}$ and $f_*^{(s,b)}$ in this equation
denote $v_*''$ and $f_*$ at the origin and the minimum of $v_*$ respectively.
Relation (\ref{fleq9}) yields a definite value only in two spacetime dimensions
where it will be used below.

In higher dimensions the unknown initial values $f_*^{(s,b)}$ enter
(\ref{fleq9}) and thus we seek a relation generalizing (\ref{fin5})
which unambiguously fixed $\xi_*$.
Comparing the second order terms in an expansion about the origin
or minimum  of (\ref{wfn1}) yields the relation
\begin{widetext}
\bea
\left(\xi_*^{(s,b)}\Big(1-\frac{\eta_*}{d+2}\Big) -
\frac{d}{24}\left(1+v_*''^{\,(s,b)}\right)\Big(1-\frac{\eta_*}{d} \Big)\right) v_*''''^{\,(s,b)} =\frac{1+v_*''^{(s,b)})}{2}\Big(1-\frac{\eta_*}{d+2}\Big)
f_*''''^{\,(s,b)}
 - \frac{\eta_*\,\xi_*^{(s,b)}}{2\mu_d}\left(1+v_*''^{\,(s,b)}\right)^3, \nn\\
\label{wfn5}
\eea
\end{widetext}
with $v''''_*$ and $f''''_*$ evaluated at the origin or the minimum.
Truncating $f_*$ to \emph{polynomials of degree two} yields
\bea
&& \left(\xi_*^{(s,b)}\Big(1-\frac{\eta_*}{d+2}\Big) -
\frac{d}{14}\left(1+v_*''^{\,(s,b)}\right)\left(1-\frac{\eta_*}{d}\right)\right)
v_*''''^{\,(s,b)}\nn \\ &&\qquad\qquad =- \frac{\eta_*\,\xi_*^{(s,b)}}{2\mu_d}
\left(1+v_*''^{\,(s,b)}\right)^3\,,\label{fleq11}
\eea
which does not depend on an unspecified initial value for $f_*$.
In numerical investigations it maybe advantageous to express
the fourth derivative of $v_*$ at the origin or minimum
by derivatives of lower order via the fixed point equation
of $v_*$.
To close the system of equations we need an equation for the
anomalous dimension. This will be discussed next.

\subsection{Flow equation for $Z_k$}

To find an equation for the anomalous dimension one must admit
an inhomogeneous field in the right hand side of the flow equation.
We have argued earlier that in the given truncation and for an even
potential a wave function renormalization only arises in a phase with
broken symmetry. In the broken phase we set
$\phi_k(x)=\phi_{0,k}+\delta\phi(x)$, where $\phi_{0,k}$
is a scale dependent minimum of the effective potential, i.e. $V'_k(\phi_{0,k})=0$.
Although a better choice would be to take the (in general inhomogeneous)
minimum of $V_k(\phi)+F_k(\phi)R$, in the following we shall take the minimum
of $V_k$ and this is justified for $\left|F_k R\right|\ll \left|V_k\right|$.
The only term in the series (\ref{act29}) which produces
a term proportional to $\delta\phi\varDelta_g \delta\phi$
is the one with $m=2$.
The effect of the nonminimal term on the position of the minima
can be taken into account perturbatively \cite{sponta}, but this issue is beyond the scope of the present work.
Since only the part $V''_k/Z_k$ of $\Sigma_k$ contributes
to the running of $Z_k$ it is sufficient to consider the first term
on the right hand side of
\bea
&& \ha\tr \left(Q_{k,1}\Sigma_k \frac{1}{P_k}\Si_k \frac{1}{P_k}\right) \nn \\
&& \qquad\qquad = \frac{1}{2Z_k^2}\tr\left(Q_{k,2}V''_k\frac{1}{P_k}V''_k\right)+O(R)\,.
\label{fleq1}
\eea
When we expand about the minimum of $V_k$
then $V_k''(0)$ in (\ref{act23})
is replaced by $V_k''(\phi_{0,k})$.
To project the first term on the \emph{r.h.s.}
onto $\int\sqrt{g}\,\phi (-\varDelta_g)\phi$ we note
that its dependence on the spacetime geometry only enters via the
covariant Laplacian in $Q_{k,2}$ and $P_k$. Thus we may take the
result in flat space time \cite{ballhausen,FRG-review,Wipf} and
just replace the Laplacian by the covariant Laplacian,
in accordance to what has been explained also in the Introduction.
With
\begin{equation}
V_k''(\phi_k)\,=\,V_k''(\phi_{0,k})
+V'''_k(\phi_{0,k})\delta\phi\,+\,\dots\label{[fleq2}
\end{equation}
 one obtains
\bea
&& \frac{V_k'''^2(\phi_{0,k})}{Z_k^2}\tr\left(Q_{k,2}
\delta\phi\frac{1}{P_k}\delta\phi\right) \label{fleq3} \\
&&\qquad = \mu_d k^{d+2}\frac{V_k'''^{\,2}(\phi_{0,k})/Z_k^2}
{(k^2+V''_k(\phi_{0,k})/Z_k)^4} \int\sqrt{g}\,\delta\phi \varDelta_g\delta\phi+\dots\,, \nn
\eea
where the dotted terms do not contribute to the running
of $Z_k$. Comparing with (\ref{act7}) finally yields
\begin{equation}
\eta_k\equiv -\frac{\partial_t Z_k}{Z_k}=\mu_d k^{d+2}
\frac{V_k'''^{\,2}(\phi_{0,k})/Z_k^3}{(k^2+V''_k(\phi_{0,k})/Z_k)^4}\,.
\label{fleq5}
\end{equation}
In terms of the dimensionless quantities in (\ref{blpa3}) this
equation reads
\begin{equation}
 \eta_k=\mu_d \frac{v_k'''^{\,2}(\chi_{0,k})}{(1+v''_k(\chi_{0,k}))^4}\,,
 \end{equation}
such that
\begin{equation}
 \eta_*=\mu_d \frac{v_*'''^{\,2}(\chi_{0*})}{\big(1+v''_*(\chi_{0*})\big)^4}\,,
 \label{fleq7}
\end{equation}
and it has been studied in detail in \cite{codello2,Hellwig:2015woa}. The last expression
yields the anomalous dimension at criticality which enters the
expression (\ref{wfn5}) and (\ref{fleq11}) for the non-minimal
couplings to the Ricci scalar.

\subsection{Numerical evaluation of $\xi_*$ in LPA'}\label{sec:numeric}
In our numerical studies we followed \cite{codello2,Hellwig:2015woa}
and first solved the fixed point equation (\ref{blpa13}) for $v_*$
with an educated first guess for $\eta_*$. With the shooting method
we determined for this $\eta_*$ the (approximate) value
of $m_*^2$ for which the differential equation admits a global solution.
From the global solution we extracted the corresponding
value of $\eta_*$ according to (\ref{fleq11}). Then we used
this value as improved guess for the shooting method. This
process is repeated until the values of $\eta_*$ converge
and one obtains a self-consistent solution of the flow equation
and the equation determining $\eta_*$. Then one calculates the
value of $\xi_*$ from this self-consistent solution.

\paragraph{Two dimensions:} We analyze the fixed point
corresponding to the critical Ising model coupled non-minimally
to gravity. From the known values of $\eta_*$ and $v''(0)$ in the Ising
and Tri-Ising class \cite{Hellwig:2015woa} we calculated $v''(\chi_{0*})$
and the corresponding values $\xi_*$ from (\ref{fleq9}):
\bea
\xi_*^{(s)}&=&\frac{1}{6}(1+v_*''\big(0)\big)\frac{1-\eta_*/2}{1-\eta_*/4}\,,\nn \\
\xi_*^{(b)}&=&\frac{1}{6}(1+v_*''\big(\chi_{0*})\big)\frac{1-\eta_*/2}{
1-\eta_*/4}\,.\label{fixedchiaw}
\eea
The results for the Ising and Tri-Ising classes are listed in
Table \ref{table:chi}.

\begin{table*}
\begin{tabular}{l|ccccccc}<
universality class&$\eta_*$&$v_*''(0)$&$v_*''(\chi_{0*})$& $\xi_*^{(s)}$&$\xi_*^{(b)}$
&$\xi_\text{class}$\\ \hline
$d=2$ Ising class LPA'&$0.4364$&$-0.3583$&$1.2489$&$0.0939$&$0.0182$&$0.0000$\\
$d=2$ Tri-Ising class LPA'&$0.3119$&$+0.2597$&$0.7534$&$0.1922$&$0.0078$&$0.0000$\\
$d=3$ Wilson-Fisher LPA&$0.0000$ &$-0.1859$ &$0.4571$ &$0.1018$&$0.1821$&$0.1250$\\
$d=3$ Wilson-Fisher LPA'&$0.1120$ &$-0.1356$ &$0.3093$ &$0.0895$&$0.1302$&$0.1250$
\end{tabular}
\caption{The anomalous dimensions, second derivative of the fixed point potentials
at the origin and the minimum and the non-minimal couplings at criticality
defined at the origin and at minimum of the scaling potential.
The last column contains the classical conformal couplings.}\label{table:chi}
\end{table*}

\paragraph{Three dimensions:} Here we assume the truncation in which
$f_*(\chi)$ is a polynomial of degree $2$, such that we may
use (\ref{fleq11}) to calculate $\xi_*$ for
fluctuations about the origin and about $\chi_{0*}$.
With the known values $\eta_*$ and $v''(0)$ from \cite{Hellwig:2015woa}
we solved the fixed point equations numerically and extracted
$v''(\chi_{0*})$ and $v''''_*(\chi_{0*})$. These then yield the
values of the minimal couplings given in Table \ref{table:chi}.
In the same Table we also included the corresponding values
for $\xi_*$ calculated in the LPA approximation with $\eta_*=0$.
In three dimensions the classical conformal coupling $\xi_\mathrm{class}=0.125$ lies
between the values extracted for fluctuations about the origin
and about the minimum of the scaling potential. This holds true in
the LPA and in the LPA' approximations.


\section{Universality and perturbation theory in $d=4-\epsilon$}\label{sec:wfr}

In this section we concentrate on the scheme-independent (universal) contribution to the flow of \eqref{act3}
with field dependent potential $V_k(\phi)$, non-minimal coupling $F_k(\phi)$ and wave function renormalization $Z_k(\phi)$.
These contributions correspond to the RG flow induced by the subtraction of the $1/\epsilon$ poles
of dimensional regularization ($\overline{\rm MS}$ scheme) below the upper critical dimension $d=4$ of a $\phi^4$ model
which is non-minimally coupled to a curved geometry.

We study the leading universal contributions in the $\epsilon$-expansion
using the approach introduced by O'Dwyer and Osborn in \cite{ODwyer:2007brp}
which was later further refined and named functional \emph{perturbative} RG in \cite{Codello:2017hhh}.
The functional perturbative flow is fully equivalent to the flow induced
by standard coupling's perturbation theory with minimal subtraction in the $\epsilon$-expansion.
In fact, all perturbative beta functionals can be derived by means of standard renormalization of the same Feynman diagrams
which in the standard approach renormalize the coupling and generate anomalous operators' scaling dimensions.

In the present work we find it more instructive
to detect which contributions are universal by extracting the
logarithmic scaling terms of the non-perturbative flow of
the full system $V_k(\phi)$, $F_k(\phi)$ and $Z_k(\phi)$.
A complete representation of the flow of this system for arbitrary cutoff can be found
in appendix \ref{app:wfr}. In the same appendix we also briefly explain which
techniques are used to extract the universal contributions and further elaborate
on other universality classes coupled to a curved geometry.

The leading universal part that is extracted from the non-perturbative RG flow in $d=4$
at the second order of the derivative expansion given in appendix \ref{app:wfr} is
\begin{align}\label{perturbative-vfz-flow-dimful}
\begin{split}
\partial_t V_k
&=
\frac{1}{(4\pi)^2} \frac{(V_k'')^2}{2Z^2_k}\,,\quad
\partial_t F_k
=
-\frac{1}{(4\pi)^2} \frac{V_k''}{Z_k}\left( \frac{1}{6} -  \frac{F_k''}{Z_k}\right)\,,
\\
\partial_t Z_k
&=
\frac{1}{(4\pi)^2} \frac{1}{Z^2_k}\left(Z_k'' V_k''+Z_k' V_k'''
\right)\,.
\end{split}
\end{align}
In the limit of small deformations of $Z_k(\phi)$ around $Z_k(\phi)=1$,
this flow can be checked against the flat-space case obtained in \cite{ODwyer:2007brp,Codello:2017hhh}.
Specifically, the flow of $V_k(\phi)$ entails the one-loop leading renormalization of the $\phi^4$ or Ising's universality class.
Furthermore, the flow of the wave function can also be checked against the results of the derivative expansion,
which also appear in \cite{ODwyer:2007brp,Codello:2017hhh}.

Let us denote the constant part of the wavefunction as $Z_{k,0}=Z_k(0)$,
which we now decorate with an additional label to distinguish it
from the full field-dependent $Z_{k}=Z_k(\phi)$.
As in Eq.~\eqref{blpa3} we define the dimensionless
field $\phi=k^{d/2-1}Z_{k,0}^{-1/2}\chi$, the dimensionless functions $v_k$ and $f_k$ and,
in addition, the dimensionless wave function renormalization
\begin{equation}
z_k(\chi) = Z_{k,0}^{-1} Z_k(\phi)\label{pertaw1}
\end{equation}
in $d=4-\epsilon$.
The rescaling of $Z_k(\phi)$ ensures the boundary condition $z_k(0)=1$.
The perturbative RG flow of these functions is
\bea
\partial_t v_k
&=&
-4 v_k + \chi v_k' +\epsilon \left(v_k -\frac{1}{2}\chi  v_k'\right)\nn \\
&& +\frac{1}{2}\eta_k
\chi v_k' +\frac{1}{(4\pi)^2} \frac{1}{2} \frac{(v_k'')^2}{z^2_k}\,,
\nn \\
\partial_t f_k
&=&
-2 f_k + \chi f'_k +\epsilon \left(f_k -\frac{1}{2}\chi  f'_k\right) \nn \\
&& +\frac{1}{2}\eta_k \chi f'_k-\frac{1}{(4\pi)^2} \frac{v_k''}{z_k} \left( \frac{1}{6} -  \frac{f_k''}{z_k}\right)\,,
\nn \\
\partial_t z_k
&=&
\eta_k z_k + \chi z'_k -\frac{\epsilon}{2}\chi z_k' +\frac{1}{2}\eta_k\chi z_k'
\nn\\
&& +\frac{1}{(4\pi)^2} \frac{1}{z_k^2}\left(z_k'' v_k''+z_k' v_k'''
\right)\,.
\label{perturbative-vfz-flow}
\eea
The anomalous dimension $\eta_k$ can be determined enforcing
the boundary condition $z_k(0)=1$,
but it is nonzero only at
two-loops \cite{ODwyer:2007brp,Codello:2017hhh},
thus it yields a correction of order $\epsilon^2$, as
expected from standard perturbation theory.
Since our results are limited to the leading order of the $\epsilon$-expansion,
we shall neglect it for the remainder of this section.

The perturbative setting simplifies the study of $k$-independent solutions of \eqref{perturbative-vfz-flow} considerably.
As expected, we find two interesting fixed points:
The non-trivial fixed point is
\begin{equation}\label{perturbative-vfz-fp}
 v_*(\chi) = \frac{1}{(4\pi)^2} \, \frac{\epsilon}{3}\,\frac{\chi^4}{4!}\,, \quad f_*(\chi) =  \frac{1}{6}\frac{\chi^2}{2!}\,, \quad z_*(\chi) = 1\,,
\end{equation}
while the generalization of the Gaussian fixed point
\begin{equation}
 v_*(\chi) = 0\,, \quad
 f_*(\chi) =  \frac{1}{6} \frac{\chi^2}{2!}\,, \quad
 z_*(\chi) = 1\,.
\end{equation}
In both cases the non-minimal coupling $\xi$ takes the expected value
$1/6$ in $d=4$.
It is interesting that the nontrivial fixed point does not exhibit the expected analytic continuation of the formula
$\xi=(d-2)/(4d-4)=1/6-\epsilon/36$ in $d=4-\epsilon$,
which makes it more difficult to interpret it as the perturbative analog of \eqref{afp34}.\footnote{
One possible point of view to understand this fact goes as follows:
Strictly speaking, the standard $\epsilon$-expansion uses the renormalization group to
trade a scaling limit in the critical coupling(s) at the fixed point for a perturbative expansion in the parameter $\epsilon$.
In this sense, all critical properties at the non-trivial fixed point in $d=4-\epsilon$, including in particular fixed points and critical exponents,
can be understood as being built from assembling data from the Gaussian theory in $d=4$.
Differently from what happens for the Wetterich's RG flow of the previous sections,
we could argue that the dimensionally regulated theory is thus never genuinely in a dimension smaller than four.
We understand, however, that this argument might not find full consensus;
for a rather different point of view on the topic and for an especially careful discussion
on how to correctly analytically continue in $d$ we suggest reading \cite{Hogervorst:2015akt}.
}
Written in this form, this result is scheme independent and therefore fully independent of any cutoff choice that was made throughout the rest of this paper, thus intuitively we could think of \eqref{afp17} as barring some explicit cutoff dependence which is ignored by the perturbative analysis.


The scaling analysis of \eqref{perturbative-vfz-flow} around the nontrivial fixed point \eqref{perturbative-vfz-fp} is also very simple.
In $d=4-\epsilon$, for arbitrarily small $\epsilon$, the mixing of the operators is selected by the mass dimension.
More precisely, we can parametrize an arbitrary deformation of the fixed point solution as
\bea
&& v_k(\chi)=v_*(\chi) + \frac{\lambda_n}{n!}\chi^n \,, \qquad
 z_k(\chi) = 1+ \frac{\zeta_{n-4}}{(n-4)!}\chi^{n-4}\,, \nn \\
 && \qquad\qquad
 f_k(\chi)= f_*(\chi) + \frac{\xi_{n-2}}{(n-2)!}\chi^{n-2}\,,
\eea
for a given natural number $n$.
The implicit condition is that only polynomial interactions are allowed $\lambda_m =\xi_m = \zeta_m=0$ if $m$ is a negative number \cite{Bridle:2016nsu}.
This implies, for example, that the first two monomials of $v_k$ cannot mix, and that the first nontrivial mixing occurs between $\phi^2$ and $R$.
This pattern continues up to the point in which all functions are mixed together starting with the (almost) marginal operators $\phi^4$, $\phi^2 R$ and $(\partial\phi)^2$.
The stability matrix in the basis $\left(\lambda_n,\xi_{n-2},\zeta_{n-4}\right)$
must be diagonalized at the fixed point \eqref{perturbative-vfz-fp}.
The negative of the eigenvalues of the stability matrix are the spectrum of scaling (critical) exponents $\theta_{n,i}$ of the theory.
We label three sets of eigenvalues
\begin{align}
\begin{split}
 &\theta_{n,1} = 4 - n -\frac{1}{6} \left(6-4n+n^2\right) \epsilon   \qquad {\rm for }\quad n\geq 0\,,\\
 &\theta_{n,2} = 4 - n -\frac{1}{6} \left(18-8n+n^2\right) \epsilon
 \hskip6mm {\rm for }\quad n\geq 2\,,\\
 &\theta_{n,3} =-\frac{1}{6} \left(n-4\right) \left( 6+(n-6)\epsilon  \right)  \qquad {\rm for }\quad n\geq 4\,.
\end{split}
\end{align}
For almost all values of $n$ the degeneracy of the critical exponents is lifted and, following the standard arguments of perturbation theory,
we could interpret the operators corresponding to the above three sets
as (normal ordered) generalizations of $\phi^n$, $\phi^{n-1} R$ and $\phi^{n-4}(\partial\phi)^2$ respectively.
In the flat-space limit, these results agree with those on the renormalization of the composite operators of the form $\phi^n$, which are known by several means
(see for example \cite{ODwyer:2007brp} and references therein).

One can also see that the use of the invariance under reparametrizations of the wavefuction has the effect that there is
exactly one marginal operator $\theta_{4,3}=0$, roughly corresponding to the kinetic term \cite{Osborn:2011kw}.
The eigenvalues of the stability matrix reveal some mixing among the considered operators.
For $n\geq 6$ the operators $\phi^n$, $\phi^{n-1} R$ and $\phi^{n-4}(\partial\phi)^2$ begin mixing with higher derivative operators.
This includes in particular those with four derivatives as discussed in \cite{Codello:2017hhh}.
We recommend \cite{Pagani:2016pad,Pagani:2017tdr} for more details on the renormalization of composite operators in the functional approach
and \cite{Pagani:2016dof} for very non-trivial applications of those results.

\section{Summary and Conclusions}
\label{S5}

We have discussed and explored functional renormalization group
(FRG) equations for the non-minimal coupling $F(\phi)R$ of a quantized
scalar field to a classical background geometry with Ricci scalar $R$.
We showed that -- similarly as in standard perturbation theory -- the
couplings in the matter sector enter the flow equation for the
scale dependent non-minimal coupling function $F_k(\phi)$ but not vice-versa.
The flow of the effective potential and field-dependent wave function
renormalization are independent of $F_k$.
In all truncations and dimensions considered the function $F_k$ fulfills
an inhomogeneous linear differential equation with coefficient functions
depending on the scale dependent effective potential $V_k$.  It
is remarkable that the $\be$-function for the dimensionless non-minimal
coupling function $f_k$ and the corresponding non-minimal coupling
$\xi_k=f_k''(0)$ is reproducing further important features of the
standard perturbative RG, which can be observed beyond one-loop order. In particular,
the FRG-based $\be$-function in $d$ dimensions
\begin{widetext}
\begin{align}
\begin{split}
\beta(\xi)=\partial_t\xi_k&=
2\mu_d\frac{v''''_k(0)}{(1+m_k^2)^3}\left(\xi_k\left(1-\frac{\eta_k}{d+2}\right)
-(1+m_k^2)\frac{d+\eta_k}{24}\right)-\mu_d\frac{f_k''''(0)}{(1+m_k^2)^2}\left(1-\frac{\eta_k}{d+2}\right),
\\ \mtxt{with}m_k^2=v_k''(0)\,,
\end{split}\label{concl3}
\end{align}
\end{widetext}
following from the flow equation (\ref{afp47}) in LPA',
does not necessarily lead to a conformal fixed point at $\xi_*=1/6$ in
four dimensions, as predicted by one-loop perturbation theory \cite{book}.
In addition, at a Gaussian fixed point with vanishing $v_*''''(0)$
we necessarily have $f_*''''(0)=0$. In LPA' and the symmetric phase
we do not observe a renormalization of the wave function. This mirrors the
same property in flat spacetime. On the other hand, in the broken phase a
non-zero wave function renormalization changes the fixed point solutions for
the non-minimal coupling function $f_*$ and the corresponding
non-minimal coupling $\xi_*$, exactly as we have described in the introduction on general grounds.
Finally, Eq.~\ref{concl3} show the IR decoupling, that was
described in the momentum-subtraction scheme of renormalization in curved
space \cite{BuGui}.

In two dimensions the equations and solutions simplify considerably.
Both in LPA and LPA' we could solve the fixed point equation for
$f_*$ analytically in terms of the fixed point potential $v_*$.
In both truncations there is an unambiguous prediction for the
non-minimal coupling, given in \eqref{afp47} and \eqref{fixedchiaw},
respectively. In LPA' one recovers all minimal models
in the Landau classification of two-dimensional conformal field
theories. From numerical solutions of the flow equation for $v_*$
with self-consistently determined anomalous dimensions one
can extract the non-minimal couplings $\xi_*$ at criticality
for this class of model. We presented the results in the symmetric
and broken phases both for the Ising and tri-Ising class.

For a sequence of dimensions between dimensions $3$ and $4$
we determined $\xi_*$ for the non-Gaussian fixed points.
From an interpolation of the corresponding values as a function
of the dimension $d=4-\epsilon$ we could numerically extract the $\epsilon$-expansion
of $\xi_*(\epsilon)$ in LPA. The same has been achieved
in the framework of the so-called functional perturbative RG, applied
to non-minimally coupled scalars in $d=4-\epsilon$ dimensions
with field and scale dependent wave function renormalization.
Besides the flow of the fixed point potential
and wave function renormalization we calculated the flow of the
non-minimal coupling function in order $\epsilon$ from the
FRG. The contribution of order $\epsilon$ to the non-minimal
coupling -- calculated numerically in LPA and analytically in the
functional perturbative RG -- are different. Future numerical efforts
with a less crude truncation may improve the situation. A first step
would be to recalculate the values in Table \ref{xistar}
in a truncation with wave-function renormalization and self-consistently
determined $\eta_*$.

In LPA' the non-minimal coupling function $f_k$ obeys a \emph{non-singular}
linear inhomogeneous differential equation. Thus parity-even fixed point
solutions depend only on one initial condition, say $f_*(0)$,
which is not quantized.
For $d\neq 2$ this free parameter enters the equation for $\xi_*$
and this ambiguity is apparently fixed by a suitable polynomial truncation
of $f_*$. It maybe interesting to see how the inclusion of the
purely gravitational contribution $\Gamma_k^\text{grav}[g]$ could
lift this degeneracy.

We have also discussed the universal contributions to the flow of the system
which appear as the logarithmically scaling terms of the renormalization group flow
and which are in one-to-one correspondence with the renormalization
induced by subtracting $1/\epsilon$ poles of dimensional regularization.
These contributions offer a different perspective on the results in $d=4-\epsilon$
and specifically on their interpolation from the four dimensional limit
in terms of universal contributions.
The universal results show the role that the cutoff
has in estimating the critical coupling $\xi$ and the critical properties in dimensions lower than four.
While a dependence on the cutoff function is a generally unwanted feature,
it is also true that only with the Wetterich equation and the scaling solutions' approach
one can obtain a realistic numerical estimation of the critical exponents of the scalar theory
in a dimensionality that is \emph{genuinely} lower than four.

A natural extension of the present work would be to
determine the running of $F_k$ in a non-minimal term of the form
\begin{equation}
\int\sqrt{g}\, F_k(\phi,R)\label{concl3}
\end{equation}
in the scale dependent effective action. Such a term
has been investigated in \cite{Narain:2009qa} with
the inclusion of metric fluctuations and the emphasis on the
asymptotic safety scenario. It is generated
during the FRG-flow from the ultraviolet to the infrared and has
been considered in studies of Higgs inflation (see, for example, \cite{defelice}).
More demanding and maybe even more interesting would be the
calculation of the dominant non-local contributions to $\Gamma_k^\mathrm{grav}$
within the FRG-approach \cite{Codello:2015oqa,Rachwal:2016vte}.

\section*{Acknowledgements}
We thank Tobias Hellwig and Benjamin Knorr for fruitful discussions
and Roberto Percacci for comments on an earlier version of the draft.
I.\ Shapiro and B.\ Merzlikin are grateful to the Theoretisch-Physikalisches-Institut
of the Friedrich-Schiller-Universit{\"a}t in Jena for
warm hospitality and support.
I.\ Shapiro is also grateful to CNPq, FAPEMIG and ICTP for partial support during the development of his work.
A.\ Wipf thanks the Departamento de F\'{i}sica of
the Universidade Federal de Juiz
de Fora for hospitality, where part
of the results in this paper have been derived.
A.\ Wipf acknowledges support from the DFG under grant no.\ Wi777/11-1.

\appendix

\section{Field-dependent wave function renormalization}\label{app:wfr}

In this appendix we present the integral form of the non-perturbative RG flow of the effective action that includes
a field dependent wave function renormalization $Z_k(\phi)$ as in \eqref{act3}.
The field dependence in the coefficient of the kinetic term makes the flow considerably more complex,
which is why we provide it in the form of a momentum space integral and only discuss some of its features
with more detail.

There are two main strategies to compute the flow of the functions $V_k(\phi)$, $F_k(\phi)$ and $Z_k(\phi)$.
On the one hand, one can use the heat kernel of the Laplacian operator $\Delta_g$ to give a computable representation
of the functional trace \eqref{wetterich} as it is done in section \ref{S2} of the main text.
On the other hand, one can obtain the same RG flows by applying a vertex expansion to \eqref{wetterich}.
In the latter case, the flows of $V_k(\phi)$, $F_k(\phi)$, and $Z_k(\phi)$
are seen respectively from the zero-point function, the two-point function of the scalar field,
and the one-point function of $h_{\mu\nu}$, where $h_{\mu\nu}$ is a small deformation of the metric
around a flat Euclidean background $g_{\mu\nu}=\delta_{\mu\nu}+h_{\mu\nu}$.
In this appendix we shall present the results derived with the vertex expansion methods described in \cite{Codello:2013wxa}.
It is a rather non-trivial check that they coincide with those coming from the heat kernel
which we derive and use in the main text, especially in the case of the one-point function of $h_{\mu\nu}$.

In order to condense the notation, let us first define a modified field
dependent propagator
\begin{equation}
 {\cal G}_k \equiv {\cal G}_k (q^2; \phi) = \left(Z_k(\phi) q^2 +V_k''(\phi) + R_k(q^2)\right)^{-1}\,,
\end{equation}
which is evaluated in momentum space and at a constant field configuration $\phi$.
The modified propagator differs from the standard propagator of the field
by the presence of the cutoff kernel $R_k(\Delta_g)$, which in flat space becomes a simple function of
the momentum square $q^2$ in agreement with its covariant form \eqref{act5}.
Let us also use primes to denote the first and second derivatives of ${\cal G}_k$ with respect to
the momentum square argument
\begin{equation}
 {\cal G}_k'= \partial_{q^2} {\cal G}_k (q^2; \phi)\,, \qquad {\cal G}_k''= \partial^2_{q^2} {\cal G}_k (q^2; \phi)\,.
\end{equation}
Ideally, the cutoff kernel is assumed to be at least twice differentiable,
but meaningful formulas can be found for optimized cutoffs such as \eqref{act31} used in the main text.
For a generic cutoff function, we find the following integral representations of the flows
\begin{widetext}
\begin{align} \label{derivative-expansion1a}
\begin{split}
 \partial_t V_k
  &=
  \int_q \frac{1}{2}{\cal G}_k \partial_t {\cal R}_k\,,
  \\
 \partial_t F_k
  &=
  \int_q \Bigl\{
  \frac{d-2}{24} \frac{1}{q^2} {\cal G}_k \partial_t {\cal R}_k -\frac{1}{2} {\cal G}_k^2 \partial_t {\cal R}_k F_k''
  \Bigr\}\,,
  \\
 \partial_t Z_k
  &=
  \int_q \Bigl\{
  \Bigl[
   {\cal G}_k' {\cal G}_k^2 \partial_t {\cal R}_k + \frac{2}{d}q^2 {\cal G}_k'' {\cal G}_k^2 \partial_t {\cal R}_k
  \Bigr] (V_k''')^2
  -\frac{1}{2} {\cal G}_k^2 \partial_t {\cal R}_k Z_k''
  \\
  &\quad
  +\Bigl[
   2 {\cal G}_k^3 \partial_t {\cal R}_k + 2\Bigl(1+\frac{2}{d}\Bigr)q^2 {\cal G}_k' {\cal G}_k^2 \partial_t {\cal R}_k + \frac{4}{d} q^4 {\cal G}_k'' {\cal G}_k^2 \partial_t {\cal R}_k
  \Bigr]Z_k' V_k'''
  \\
  &\quad
  +\Bigl[
   \Bigl(2+\frac{1}{d}\Bigr) q^2 {\cal G}_k^3 \partial_t {\cal R}_k + \Bigl(1+\frac{4}{d}\Bigr) q^4 {\cal G}_k' {\cal G}_k^2 \partial_t {\cal R}_k +\frac{2}{d} q^6 {\cal G}_k'' {\cal G}_k^2 \partial_t {\cal R}_k
  \Bigr](Z_k')^2
  \Bigr\}\,.
\end{split}
\end{align}
\end{widetext}
The momentum space measure is normalized by including all factors of $(2\pi)$ as
\begin{equation}
 \int_q = \frac{1}{(2\pi)^d}\int {\rm d}^d q = \frac{1}{(2\pi)^d}\int {\rm d}q \, q^{d-1} {\rm d}\Omega_{d-1}\,.
\end{equation}
We used rotational and translational invariance to arrange all integrands of \eqref{derivative-expansion1a} so that they are manifestly functions of $q^2$.
The angular integration is thus decoupled and one could already use the volume of the $d$-sphere
to obtain the results of the main text. More precisely
it is necessary to switch to the integration variable $z=q^2$ and use the definition of $\mu_d$ \eqref{lpa7}
\bea
 \int_q &=& \frac{1}{(2\pi)^d} \frac{2 \pi^{\frac{d}{2}}}{\Gamma(\frac{d}{2})}\int {\rm d}q \, q^{d-1}
 = \frac{1}{(4\pi)^{\frac{d}{2}}} \frac{1}{\Gamma(\frac{d}{2})} \int {\rm d}q^2 (q^2)^{\frac{d}{2}-1} \nn\\
 &=&  \frac{2\mu_d}{d} \int {\rm d}z z^{\frac{d}{2}-1}
\eea
to recover the integrations of section \ref{S2}.

Using the system \eqref{derivative-expansion1a}, it is possible to derive all the
RG flows given in the main text.
The flows in the LPA given in \eqref{lpa11a} and \eqref{lpa11b} can be obtained by setting $Z_k(\phi)=1$ and choosing the cutoff $R_k(q^2)=(k^2-q^2)\theta(k^2-q^2)$.
The flows in the LPA' given in \eqref{blpa1} can similarly be obtained by setting $Z_k(\phi)={\rm constant}$ and choosing the cutoff $R_k(q^2)=Z_k (k^2-q^2)\theta(k^2-q^2)$.
In the latter case, the flow of the wave function $Z_k$ depends only on the first two terms of $\partial_t Z_k$ in \eqref{derivative-expansion1a}
which are generally evaluated at the minimum of the potential, being it the field configuration that represents the ground state of the quantum theory.

We can also use the system \eqref{derivative-expansion1a}
to study the leading universal perturbative features of the RG flow close to some interesting upper critical dimensions\footnote{
In statistical physics, the upper critical dimension is generally defined as the highest dimensionality
in which the system has a nontrivial second order phase transition.
From the point of view of the RG, above the upper critical dimension fluctuations are weak,
and thus the phase transition is controlled by the Gaussian fixed point while the critical properties coincide with their mean field estimates.
Below the upper critical dimension the phase transition is instead controlled by a nontrivial fixed point,
and the scaling analysis receives sizeable corrections from the fluctuations of the field.
}
in the approach that goes under the name of functional \emph{perturbative} RG \cite{Codello:2017hhh}.
These universal contributions to the RG do not depend on the cutoff $R_k$ and can be either seen as coming from the subtraction
of logarithmic divergences or alternatively as the terms scaling with ``momentum to the power zero''
in the beta functions (see for example the appendix of \cite{Codello:2017hhh}).
They are completely equivalent to what one would obtain from minimal subtraction of divergences in dimensional regularization.
Given a certain value for the dimension which plays the role of \emph{upper} critical dimension,
the simplest strategy to find the monomials corresponding the the perturbative flow is to choose
a mass cutoff $R_k=k^2$ (the simplest cutoff) and determine from the non-perturbative flow which terms scale as $k^0$,
while neglecting all other relevant and irrelevant contributions.

As an illustrative example, let us derive the functional perturbative flow for the potential in $d=4$ in the LPA.
We expand Eq.~\eqref{derivative-expansion1a} in powers of $V''(\phi)$ and notice that
\begin{equation}
 \partial_t V_k = \frac{\mu_4}{4} \int {\rm d}q^2 q^2 \frac{k^2}{Z_k q^2+k^2} (V'')^2 +\dots\,,
\end{equation}
where in the dots are hidden the terms that either diverge in the limit $k\to0$ (UV irrelevant) or $k\to\infty$ (UV relevant).
By construction, standard dimensional regularization is insensitive of the same terms,
because it lacks a momentum scale that is necessary to give a nonzero value to those integrals (a role that is played by $k$ in this context).
The perturbative part of the flow can be obtained by simply eliminating all the terms hidden in the dots.
Upon elimination we find
\begin{equation}
 \partial_t V_k = \frac{1}{(4\pi)^2}\frac{1}{2}\frac{(V'')^2}{Z_k^2}\,,
\end{equation}
which coincides with the result given in section \ref{sec:wfr}.
It is very simple to follow the same strategy for the flow of the other two functions
and obtain the full system \eqref{perturbative-vfz-flow-dimful}.

The functional perturbative RG is fully equivalent to the standard
perturbation theory that is obtained by minimal subtraction of the
$\frac{1}{\epsilon}$ poles of dimensional regularization
\cite{ODwyer:2007brp,Osborn:2011kw,Osborn:2017ucf}. We discuss the
perturbative results for the full system of functions $V_k(\phi)$,
$F_k(\phi)$ and $Z_k(\phi)$ in $d=4-\epsilon$ dimensions in section
\ref{S5}. Interestingly, however, the non-perturbative flow
\eqref{derivative-expansion1a} is suitable to find the perturbative
contributions in proximity of two additional interesting upper
critical dimensions, namely $d=2$ and $d=6$. We display all further
results using the convention that renormalized fields are obtained
by rescaling the full field-dependent wavefunction $Z_k(\phi)$. This
procedure does not change the spectrum of scaling operators; for a
more detailed analysis that includes the full mixing of the
wavefunction we refer to \cite{ODwyer:2007brp}.

As shown in \cite{Codello:2017hhh}, the use of $d=2$ as upper critical dimension highlights RG equations
for the Sine-Gordon universality class. We find that leading universal contributions to the flow in $d=2$ are
\bea
 && \partial_t V_k
 = -\frac{1}{4\pi} \frac{V_k''}{Z_k}\,,
 \quad
 \partial_t F_k
 = -\frac{1}{4\pi}\frac{F_k''}{Z_k}\,,
 \quad
 \partial_t Z_k
 = -\frac{1}{4\pi}\frac{Z_k''}{Z_k}\,. \label{perturbative2d} \nn
\eea
The relation with the Sine-Gordon model is best seen in the LPA by simply switching to dimensionless renormalized variables $v_k(\chi) =k^{-2} V_k(\chi)$
at $\eta=0$ and solving the fixed point equation for $v_*(\chi)$
\begin{equation}
 v_*(\chi) = -\frac{m^2_*}{8\pi}\cos\left(\sqrt{8\pi}\chi\right)\,,
\end{equation}
which uses the boundary condition $v_*''(0)=m^2_*$ and manifestly displays the value
$\sqrt{8\pi}$ known as Coleman phase (see also the discussion of \cite{Codello:2017hhh}).
The two dimensional system thus hints at the existence of a generalization of the Sine-Gordon universality class that is coupled to a fixed geometry.
The system could in principle be used to estimate the central charge of the Sine-Gordon model upon integration of the flow \cite{Bacso:2015ixa} from UV to IR,
but the flows of all functions are decoupled.
We hope to come back on this topic in the future.

In $d=6$ the leading universal part of \eqref{derivative-expansion1a} is only slightly more involved.\footnote{Actually
the flow \eqref{derivative-expansion1a} contains several more cutoff-independent terms,
but we consistenly display only the ones that are responsible to the leading corrections in the $\epsilon$-expansion.
}
We find
\bea
  \partial_t V_k &=& -\frac{1}{(4\pi)^3} \frac{1}{6} \frac{(V_k'')^3}{Z_k^3}\,, \nn\\
 \partial_t F_k
 &=& \frac{1}{(4\pi)^3}\left( \frac{1}{6} -\frac{F_k''}{Z_k}\right)\frac{1}{2}\frac{(V_k'')^2}{Z_k^2}\,,\nn
 \\
 \partial_t Z_k
 &=& -\frac{1}{(4\pi)^3} \frac{1}{6} \frac{(V_k''')^2}{Z_k^2}
 \eea
This system can also be checked against the leading one-loop contributions in the flat space limit \cite{Codello:2017hhh,Zambelli:2016cbw}.
Upon moving to dimensionless renormalized variables $\chi,v_k,f_k$ and $z_k$ defined in \eqref{pertaw1} with $Z_{k,0}=Z_k(0)$
in $d=6-\epsilon$, we find that the system admits the non-trivial fixed point solution
for the dimensionless renormalized functions
\bea
 v_*(\chi) &=& \frac{1}{(4\pi)^{3/2}} \,\sqrt{-\frac{2\epsilon}{3}}\, \frac{\chi^3}{3!}\,, \nn \\
 f_*(\chi) &=& \frac{1}{5}\, \frac{\chi^2}{2!} \,,
 \qquad z_*(\chi)=1\,,
\eea
with anomalous dimension $\eta=-\epsilon/9$ (at criticality the model
has negative $\eta$ and does not satisfy the unitarity bound).
This fixed point corresponds to a Lee-Yang universality class \emph{minimally} coupled to the curved geometry, as seen from the non-minimal coupling $\xi_*=f_*''(0)=(d-2)/(4d-4)=1/5$ in $d=6$.
Similarly to the case $d=4$ discussed in section \ref{S5}, we do not find $\epsilon$-corrections
to the non-minimal coupling
even though one would naively expect $\xi_* =1/5-\epsilon/100$ from expanding in $d=6-\epsilon$ the $d$-dependent formula for $\xi_*$.
Notice however, that the above statement on the absence of ${\cal O}(\epsilon)$ corrections is restricted to the leading order in the $\epsilon$ expansion,
and the inclusion of the next-to-leading two loops contributions might as well result in new nontrivial contributions to $\xi_*$ at order $\epsilon^2$.


\end{document}